\documentclass[conference]{IEEEtran}
\usepackage{cite}
\usepackage{amsmath,amssymb,amsfonts}
\usepackage{graphicx}
\usepackage{textcomp}
\usepackage{xcolor}
\usepackage{url}
\usepackage{algorithm}
\usepackage{algorithmic}
\usepackage{float}

\usepackage[top=0.7in, bottom=1in, left=0.7in, right=0.7in]{geometry}
\newcommand{\E}{\mathbb{E}}
\newcommand{\C}{\mathcal{C}}
\newcommand{\Dset}{\mathcal{D}}

\begin{document}

\title{Channel-Aware Codebook Design for Task-Oriented Token Communication Over Wireless Networks}

\author{
\IEEEauthorblockN{Chenliang Wu, Zhicheng Liu, Zhouxiang Zhao, Ruopeng Xu, Zhaohui Yang, and Zhaoyang Zhang}
\IEEEauthorblockA{
        College of Information Science and Electronic Engineering, Zhejiang University, Hangzhou 310027, China\\
        E-mail: \{chenliangwu, liu\_zc, zhouxiangzhao, ruopengxu, yang\_zhaohui, ning\_ming\}@zju.edu.cn}

\vspace{-1em}
}

\maketitle

\begin{abstract}
This paper investigates the adaptive codebook design for task-oriented token communication over wireless networks with dynamic channels. This
paper proposes the channel-aware codebook (CAC), a joint channel- and reconstruction-driven framework that reorders the labels of a frozen 14-bit VQGAN codebook. In the considered framework, a balanced geometric partition forms prefix groups, a prefix bijection
places the groups on the label hypercube, and bijective suffix assignments place the codewords within each group. The problem is formulated to minimize the  expected full-image perceptual distortion of the frozen decoder under a binary symmetric channel, averaged over a design set of bit error rates (BERs) and retaining simultaneous token errors. An offline algorithm alternates prefix and suffix exchanges and screens and confirms candidates on paired batches under a fixed decoder-evaluation budget.
The assignment preserves the error-free reconstruction and the payload and deploys as static lookup tables without retraining. Under the binary symmetric channel at BER $0.01$, it reduces LPIPS by $12.3\%$ and improves COCO AP by $14.1\%$ ($1.72$ points) relative to the native labels, and it lowers LPIPS by a further $4.5\%$ relative to a related assignment adaptation at the same payload.
\end{abstract}


\section{Introduction}
Artificial intelligence (AI) is moving from independent inference to perception and action across connected devices. Embodied agents and emerging sixth generation (6G) services exchange information to complete shared tasks~\cite{xu2025energy,zhao2026fluidantennasystemsisac}. This shift motivates semantic communication and communication designed for specific tasks, where system design focuses on the usefulness of the received information~\cite{gunduz2023,11006980}. Tokens connect learned AI representations with digital communication systems. In visual token communication, each token identifies a codeword in a learned image representation. Transmitting these indices instead of raw pixels provides a compact digital payload that a compatible model can reconstruct or use for further processing. Token communication can therefore support information exchange among intelligent devices over wireless networks.

A typical visual token communication system extracts image features, quantizes them into codewords, transmits their binary labels, and reconstructs the image or performs a task at the receiver. Communication designed for a specific task focuses on the usefulness of the received information. Learned joint source and channel coding extends image communication to the physical layer~\cite{kurka2020}. VQ-VAE learns a discrete codebook together with an encoder and a decoder~\cite{vandenoord2017}. VQGAN adds perceptual and adversarial training to support image synthesis at high resolutions~\cite{esser2021}. For label assignment, Kutay and Yener use the geometry of pretrained embeddings to train a neural labeling model and then obtain a bijective assignment~\cite{kutay2025}. TONIC combines unequal error protection based on token utility, confidence gating, and token completion using a Transformer~\cite{liu2026tonic}. These methods improve different parts of the system, including image representation, token transmission, and recovery from token errors.

The mapping between a visual codeword and its transmitted binary label determines which codeword the receiver selects after a channel error. Classical vector quantization represents a source using a finite codebook~\cite{gersho1992}. Index assignment then chooses the labels to reduce expected distortion under the channel model~\cite{farvardin1990,talbot1997,knagenhjelm1996}. For pretrained visual tokenizers, however, distances between latent vectors do not fully describe the perceptual effects of simultaneous token errors after nonlinear image decoding. This leaves a need for label assignments that account for the image distortion caused by channel errors in a large, fixed visual codebook. We address this problem with the channel aware codebook (CAC). CAC optimizes the label assignment offline using channel samples and the fixed visual decoder. During transmission, it changes only the lookup mapping, preserves the number of bits per token, and requires no retraining of the visual backbone.

The main contributions of this paper are summarized below.
\begin{enumerate}
\item We propose CAC, a framework that reorders the labels of a fixed visual codebook using both the channel model and reconstruction quality. The objective is the expected perceptual distortion of complete images reconstructed by the fixed decoder under channel errors. The design directly uses channel samples and decoded images rather than an approximate distortion measure. It preserves the payload size and the reconstruction obtained without channel errors.
\item We develop an offline algorithm to optimize this objective. A balanced geometric partition divides the codewords into groups of equal size. A prefix bijection assigns labels to groups, and suffix bijections assign labels to codewords within each group. Candidate exchanges are screened and confirmed using paired evaluations within a fixed budget of decoder calls. The final checkpoint is selected on a development set.
\item We evaluate CAC using a VQGAN codebook with 14 bits per token on 2048 images from COCO val2017. At a BER of $0.01$, CAC reduces LPIPS by $12.3\%$ and increases COCO AP by $14.1\%$, or $1.72$ points, compared with the native labels. It also achieves lower LPIPS than an adapted assignment method trained using embedding geometry, with the same payload size.
\end{enumerate}

\section{System Model and Problem Formulation}
\label{sec:model}
A token communication system transmits the indices produced by a learned tokenizer instead of raw samples~\cite{gunduz2023}. At the transmitter, an encoder maps the input to a grid of latent vectors. A vector quantizer replaces each vector with its nearest codeword, and the corresponding index is converted into a binary label. After channel protection or modulation, the labels are transmitted over the wireless channel. The receiver uses the recovered indices to select vectors from the shared codebook. A decoder then reconstructs the signal, which can be used by a task model~\cite{kurka2020}. The codebook is learned together with the encoder and decoder~\cite{vandenoord2017}. Perceptual and adversarial training support reconstruction at high resolutions~\cite{esser2021}. We consider a system with a pretrained VQGAN encoder $E$, vector quantizer $Q$, decoder $D$, and codebook given by
\begin{equation}
 \C=\{\mathbf e_i\}_{i=0}^{K-1},\qquad
 \mathbf e_i\in\mathbb R^d,\quad K=2^B.
\end{equation}
For an image $x$, the encoder and quantizer produce a sequence of $L$ original codeword indices. The corresponding reconstruction without channel errors is $x_0$.
\begin{equation}
 \mathbf z(x)=Q(E(x))=(z_1,\ldots,z_L),\qquad
 x_0=D(\C[\mathbf z(x)]),
 \label{eq:visual}
\end{equation}
Here, $\C[\mathbf z]$ denotes the spatial grid of selected codeword vectors. The visual backbone and all codeword vectors remain fixed during assignment training. We use $K=16384$, $d=256$, and $B=14$. The model also applies to any codebook whose size is a power of two.

A bijection $\pi:\{0,\ldots,K-1\}\rightarrow\{0,1\}^B$ maps each original codeword index to a binary transmission label of length $B$. We call this mapping the index assignment or index reordering. The transmitter applies $\pi$ after quantization. For a received label $\hat{\mathbf b}_\ell$, the receiver recovers the index as $\hat z_\ell=\pi^{-1}(\hat{\mathbf b}_\ell)$ and reconstructs the image using
\begin{equation}
 \hat x_\pi=D(\C[\hat{\mathbf z}]),\qquad
 \C_{\mathrm{CAC}}^{\pi}[\mathbf b]=\mathbf e_{\pi^{-1}(\mathbf b)}.
 \label{eq:wirelesscodebook}
\end{equation}
Here, $\C_{\mathrm{CAC}}^{\pi}$ is the codebook indexed by transmission labels. The transmitter and receiver share the same assignment. The payload contains $LB$ bits. Since the mapping is bijective, $\C_{\mathrm{CAC}}^{\pi}[\pi(i)]=\mathbf e_i$, so the reconstruction remains $x_0$ when no channel errors occur.

\begin{figure}[t]
\centering
\includegraphics[width=\columnwidth]{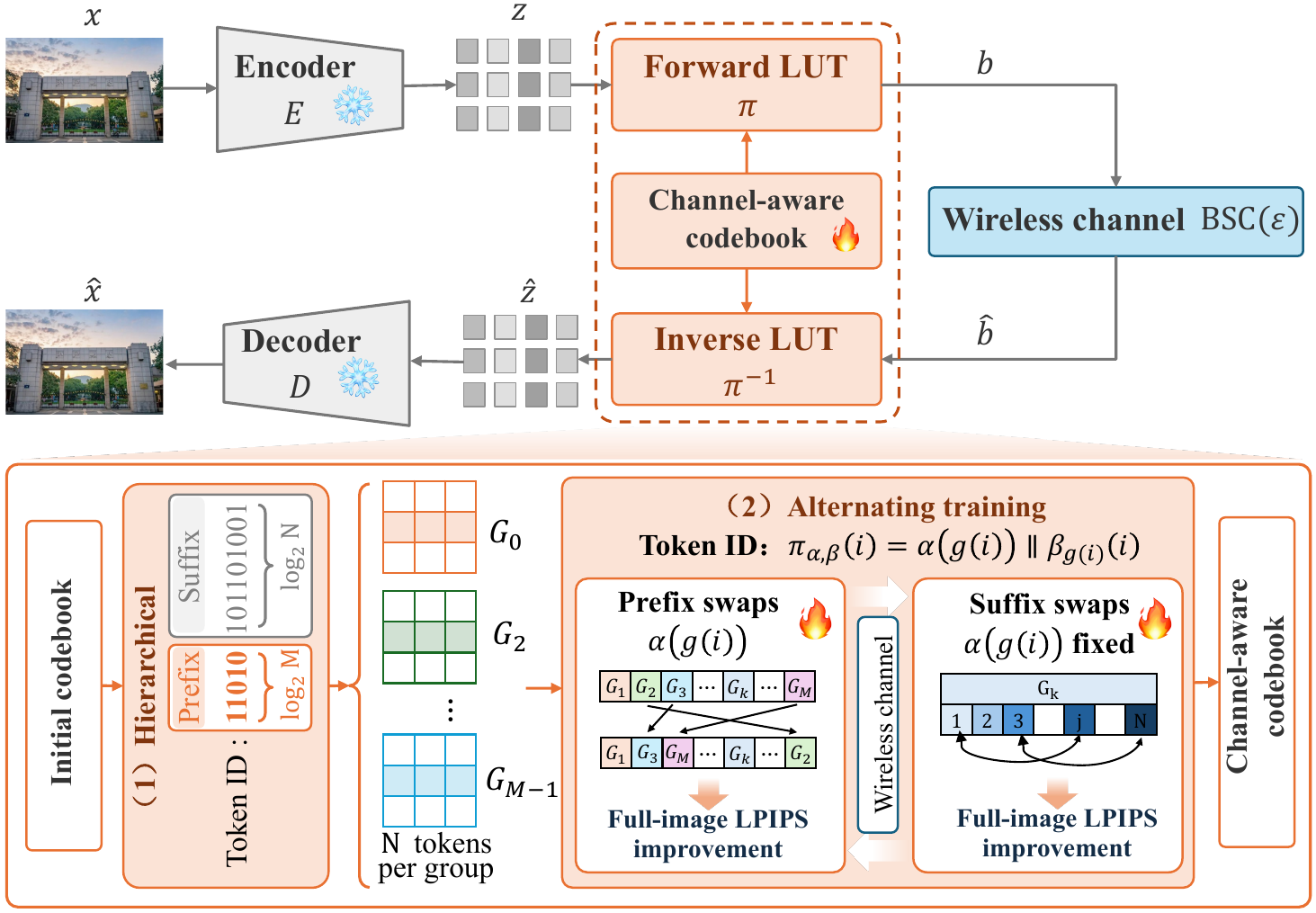}
\caption{Channel aware codebook design with offline index reordering and online token transmission using lookup tables.}
\label{fig:system}
\vspace{-1em}
\end{figure}

Fig.~\ref{fig:system} shows the proposed method. It keeps the tokenizer fixed and changes only the mapping between codeword indices and transmission labels. A balanced geometric partition first divides the $K$ codewords into $M$ groups of equal size, placing nearby codewords in the same group. Codewords in each group share a label prefix. The algorithm then alternates between exchanges of group prefixes and exchanges of suffixes within each group. It accepts an exchange only when paired evaluations indicate a reduction in the expected image distortion under the channel model in Section~\ref{sec:channel}. The selected assignment is stored as forward and inverse lookup tables. Online transmission therefore requires one lookup at the transmitter and one at the receiver for each token. Sections~\ref{sec:hier}, \ref{sec:channel}, and \ref{sec:objective} describe the hierarchical assignment, the channel and distortion models, and the optimization objective, respectively. Section~\ref{sec:algorithm} presents the offline search algorithm.

\subsection{Hierarchical Index Reordering Model}
\label{sec:hier}
Let $K=MN$, where $M=2^p$ is the number of groups and $N=2^q$ is the number of codewords in each group. Each group has $N$ suffix positions, and $p+q=B$. The codebook is divided into disjoint groups of equal size according to
\begin{equation}
 \C=\bigcup_{g=0}^{M-1}\C_g,\qquad
 \C_g\cap\C_h=\varnothing\;(g\neq h),\qquad |\C_g|=N.
 \label{eq:partition}
\end{equation}
Let $g(i)$ denote the group containing codeword $i$. To place nearby codewords in the same group, we use the following initialization criterion.
\begin{equation}
 \min_{\{\C_g\}}\sum_{g=0}^{M-1}\sum_{i:\mathbf e_i\in\C_g}
 \|\mathbf e_i-\boldsymbol\mu_g\|_2^2,
 \quad |\C_g|=N,
 \label{eq:geometry}
\end{equation}
Here, $\boldsymbol\mu_g$ is the centroid of group $g$. We construct the groups by recursively splitting the codeword vectors. At each step, we project the vectors in a subset onto its leading principal direction, sort the projections, and divide them into two equal subsets. Cutting the resulting tree at depth $p$ gives the groups, while the remaining branches provide the initial suffix labels. This procedure produces a valid assignment without requiring a global solution to~\eqref{eq:geometry}. Group membership remains fixed during label optimization.

The prefix mapping $\alpha$ assigns a unique prefix to each group. For each group $g$, the suffix mapping $\beta_g$ assigns a unique suffix to each codeword. Both mappings are bijections.
\begin{align}
 \alpha &: \{0,\ldots,M-1\}\rightarrow\{0,1\}^p,\\
 \beta_g &: \{i:\mathbf e_i\in\C_g\}\rightarrow\{0,1\}^q.
\end{align}
The complete label is formed by joining the group prefix and the codeword suffix.
\begin{equation}
 \pi_{\alpha,\beta}(i)=\alpha(g(i))\,\Vert\,\beta_{g(i)}(i),
 \label{eq:hierarchy}
\end{equation}
Here, $\Vert$ denotes concatenation. The mapping $\alpha$ places the groups on the prefix hypercube, and $\beta_g$ places the codewords of group $g$ on the suffix hypercube. Every integer $p$ from $1$ to $B-1$ gives a valid split with $M=2^p$ and $N=2^{B-p}$. For $B=14$, there are thirteen possible splits. Section~\ref{sec:simulations} gives the split used in our experiments. The hierarchical assignment allows any permutation of group prefixes and independent permutations of suffixes within each group. The two exchange operations described below can generate all assignments in this family. For $M,N>1$, however, this family contains only a subset of the $K!$ possible bijections.

The two assignment levels are linked through the Hamming distance between labels.
\begin{align}
 d_H(\pi(i),\pi(j))
 &=d_H(\alpha(g(i)),\alpha(g(j)))\nonumber\\
 &\quad+d_H(\beta_{g(i)}(i),\beta_{g(j)}(j)).
 \label{eq:hamming}
\end{align}
If an error changes the prefix but leaves the suffix unchanged, the receiver selects the codeword with that suffix in another group. Suffix optimization therefore affects both substitutions within a group and the reconstruction effects of substitutions between groups.

We alternate between two types of exchanges. A prefix exchange swaps $\alpha(g)$ and $\alpha(h)$ for two groups while keeping their suffix assignments unchanged. A suffix exchange swaps $\beta_g(i)$ and $\beta_g(j)$ for two codewords in the same group. Both operations preserve the bijection. For a source codeword $i$, we restrict suffix candidates to
\begin{equation}
 \mathcal N_g(i)=\{j:\mathbf e_j\in\C_g,\;
 1\le d_H(\beta_g(i),\beta_g(j))\le2\}.
 \label{eq:neighbors}
\end{equation}
This set contains $q+\binom{q}{2}$ candidates, compared with $N-1$ candidates when all other codewords in the group are considered. We rank these candidates by vector distance and use decoder evaluations to decide whether to accept an exchange. Prefix proposals use group geometry to prioritize group pairs. Each round first searches prefix exchanges and then suffix exchanges. Source codewords for suffix exchanges are sampled according to their occurrence frequencies. The partition defines the search structure, and the same reconstruction objective is used in both phases.

\subsection{Channel Model and Reconstruction Distortion}
\label{sec:channel}
We model errors in the detected binary labels using an independent binary symmetric channel (BSC) with bit error probability $\varepsilon$. This is a standard model for a memoryless binary link with symmetric errors~\cite{cover2006}. For each token, the received label is given by
\begin{equation}
 \hat{\mathbf b}_\ell=\pi(z_\ell)\oplus\boldsymbol\eta_\ell,
 \qquad \eta_{\ell,r}\sim\operatorname{Bernoulli}(\varepsilon),
 \label{eq:bscsample}
\end{equation}
The noise bits are independent across token positions $\ell$ and bit positions $r$. The probability that transmitted codeword $i$ is received as codeword $j$ is
\begin{equation}
 P_\varepsilon(j\mid i;\pi)
 =\varepsilon^{d_{ij}}(1-\varepsilon)^{B-d_{ij}},
 \quad d_{ij}=d_H(\pi(i),\pi(j)).
 \label{eq:bsc}
\end{equation}
The assignment determines which codeword is selected after an error. The token error probability is $1-(1-\varepsilon)^B$, regardless of the assignment. Changing the numbers of prefix and suffix bits changes how these errors are distributed between the two levels.

For a complete noise grid $\boldsymbol\eta$, the reconstructed image is
\begin{equation}
 \hat x_\pi(x,\boldsymbol\eta)
 =D\!\left(\C\!\left[
 \pi^{-1}(\pi(\mathbf z(x))\oplus\boldsymbol\eta)
 \right]\right),
 \label{eq:noisyimage}
\end{equation}
The mappings are applied separately to each token. We measure training distortion using LPIPS~\cite{zhang2018} between the original image and the complete reconstructed image.
\begin{equation}
 L_{\rm LPIPS}(x,\boldsymbol\eta;\pi)
 =\operatorname{LPIPS}(x,\hat x_\pi(x,\boldsymbol\eta)).
 \label{eq:loss}
\end{equation}
Equations~\eqref{eq:noisyimage} and~\eqref{eq:loss} account for simultaneous errors across the token grid and their combined effects through the fixed decoder. The local Hamming neighborhoods restrict only the proposed exchanges. Channel sampling still allows every possible pattern of bit errors.

For uncoded coherent BPSK over AWGN, the BER is related to the signal to noise ratio by $\varepsilon=\tfrac12\operatorname{erfc}(\sqrt{\gamma_b})$, where $\gamma_b=E_b/N_0$. The experiments evaluate reconstruction quality, detection accuracy, and comparisons with related methods over a range of BERs.

\subsection{Objective Based on Channel Errors and Reconstruction Quality}
\label{sec:objective}
Let $\mathcal E$ be a finite set of training BERs. Each BER has a nonnegative weight $w_\varepsilon$, and the weights sum to one. For a fixed geometric partition, we optimize the following objective.
\begin{subequations}\label{eq:objective}
\begin{align}
&\underset{\alpha,\{\beta_g\}}{\min}\;J_M(\alpha,\beta)
=\sum_{\varepsilon\in\mathcal E}w_\varepsilon\E_{x,\boldsymbol\eta}
\big[L_{\rm LPIPS}(x,\boldsymbol\eta;\pi_{\alpha,\beta})\big]
\tag{\ref{eq:objective}}\label{eq:risk}\\
&\text{s.t.}\quad\alpha:\{0,\ldots,M-1\}\to\{0,1\}^{p},\
\label{eq:con-alpha}\\
&\phantom{\text{s.t.}\quad}\beta_g:\{i:\mathbf e_i\in\C_g\}\to\{0,1\}^{q},\  \forall g,
\label{eq:con-beta}\\
&\phantom{\text{s.t.}\quad}\pi_{\alpha,\beta}(i)=\alpha(g(i))\,\Vert\,\beta_{g(i)}(i),\ \forall i.
\label{eq:con-pi}
\end{align}
\end{subequations}
Constraint~\eqref{eq:con-alpha} requires a bijection over the $M$ groups, so each prefix is assigned to exactly one group. Constraint~\eqref{eq:con-beta} requires a bijection within each group, so each of its $N$ suffixes is assigned to exactly one codeword. Together, these constraints preserve the complete bijection and restrict the search to the hierarchical assignments described in Section~\ref{sec:hier}. Constraint~\eqref{eq:con-pi} forms each label by joining its group prefix and codeword suffix. Thus, only the prefix and suffix assignments are optimized.

The expectation in~\eqref{eq:risk} is taken over training images $x\sim\Dset_{\rm tr}$ and noise masks drawn from the BSC at each BER. This retains the token occurrence frequencies in the training images. Since the reconstruction without channel errors does not depend on the assignment, subtracting $\E_x[\operatorname{LPIPS}(x,x_0)]$ gives an equivalent objective that measures the additional distortion caused by channel errors. Under the independent BSC model, applying the same XOR mask to every label or permuting the bit positions of every label leaves the expected distortion unchanged. However, exchanging prefix and suffix positions does not generally preserve the hierarchical constraints. Bit order therefore does not affect the distortion of a fixed assignment, while the choice of $p$ and $q$ changes the allowed assignments. At $\varepsilon=1/2$, all bijections have the same expected distortion.

We estimate the objective using Monte Carlo reconstruction. To compare the current assignment $\pi$ with a candidate $\pi'$, we use the same images and noise masks for both assignments.
\begin{equation}
\begin{split}
 \widehat{\Delta J}
 &=\sum_{\varepsilon\in\mathcal E}\frac{w_\varepsilon}{S}
 \sum_{s=1}^{S}
 \Big[L_{\rm LPIPS}(x_s,\boldsymbol\eta_{s,\varepsilon};\pi')\\
 &\qquad -L_{\rm LPIPS}(x_s,\boldsymbol\eta_{s,\varepsilon};\pi)\Big].
\end{split}
 \label{eq:paired}
\end{equation}
A small batch is used to screen candidate exchanges. Candidates that pass screening are evaluated on an independent confirmation batch before acceptance. Prefix and suffix exchanges use the same criterion to assess their effects on reconstruction quality. The search follows the decoder call budget and checkpoint selection rule in Section~\ref{sec:algorithm}. For the fixed grouping used in our experiments, checkpoints are selected using the weighted objective in~\eqref{eq:risk} on the development set.

Section~\ref{sec:algorithm} describes the offline search procedure. Section~\ref{sec:simulations} evaluates the selected assignment under the channel model above. Detection AP is used only for evaluation and is not part of the training objective.

\section{Offline Index Reordering Algorithm}
\label{sec:algorithm}

The algorithm takes the fixed encoder $E$, quantizer $Q$, decoder $D$, and codebook $\C$, together with the training and development sets $\Dset_{\rm tr}$ and $\Dset_{\rm dev}$, the partition $\{\C_g\}$, the BER set $\mathcal E$, its weights $w_\varepsilon$, and a decoder call budget $C_{\rm dec}$. It returns $\alpha$, $\{\beta_g\}$, and the forward and inverse lookup tables. The encoder, quantizer, decoder, and codebook remain unchanged. The algorithm performs a random local search over the hierarchical assignments in Section~\ref{sec:model} within the given budget. Each valid exchange is screened using paired evaluations and then checked on an independent batch. It is accepted only if it passes the confirmation test.

We build the geometric tree once. Cutting it at depth $p$ gives the groups $\C_g$. The path from the root to group $g$ gives its initial prefix $\alpha_0(g)$, and the remaining path to codeword $i$ gives its initial suffix $\beta_{0,g}(i)$. All runs therefore start from the same assignment, and group membership remains fixed. One encoder pass over $\Dset_{\rm tr}$ stores the token sequences $\mathbf z(x)$ and estimates the token frequencies $\hat f_i$. We form $\mathcal A$ by adding codewords in descending order of $\hat f_i$ until their cumulative frequency reaches or exceeds $\rho=0.9$. Ties are resolved by ascending codeword index. The confirmation rule requires a minimum number of source tokens associated with an exchange. We therefore focus suffix proposals on frequent source codewords in $\mathcal A$, while keeping all codewords in the assignment.

The search uses prefix exchanges $\alpha(g)\leftrightarrow\alpha(h)$ and suffix exchanges $\beta_g(i)\leftrightarrow\beta_g(j)$. For a prefix proposal, group $g$ is sampled with probability proportional to its frequency $\Lambda_g=\sum_{i:\mathbf e_i\in\C_g}\hat f_i$. With probability $1/2$, group $h$ is sampled uniformly from all other groups. Otherwise, it is sampled uniformly from the $\kappa=8$ groups with centroids closest to $\boldsymbol\mu_g$. For a suffix proposal, source codeword $i\in\mathcal A$ is sampled with probability $\hat f_i/\sum_{j\in\mathcal A}\hat f_j$. Candidates in $\mathcal N_g(i)$, whose suffix labels differ from that of $i$ in one or two bits, are ranked by increasing vector distance $\|\mathbf e_i-\mathbf e_j\|_2$. The algorithm tries up to the first $k=4$ candidates in this order. This limit reduces the number of decoder evaluations per source. Vector distance determines the candidate order, while decoder evaluations determine acceptance. If a candidate is rejected, the next candidate is tested. If all candidates fail, the algorithm samples another source. Each unordered pair of groups or codewords is tested at most once per round.

For a candidate exchange from $\pi$ to $\pi'$, let $\bar\delta_s$ denote the weighted difference in LPIPS for image $s$ across the training BERs, given by $\bar\delta_s=\sum_{\varepsilon}w_\varepsilon[L_{\rm LPIPS}(x_s,\boldsymbol\eta_{s,\varepsilon};\pi')-L_{\rm LPIPS}(x_s,\boldsymbol\eta_{s,\varepsilon};\pi)]$. Both assignments use the same image and noise mask at each BER. Screening uses $S=8$ images from $\Dset_{\rm tr}$. The candidate is rejected if the mean difference is nonnegative. Otherwise, an independent batch of $S'=32$ images and new noise masks is used for confirmation.
\begin{equation}
 \widehat{\Delta J}_{\rm conf}=\frac1{S'}\sum_{s=1}^{S'}\bar\delta_s,
 \qquad
 \widehat{\Delta J}_{\rm conf}+1.645\,\widehat{\mathrm{SE}}_{\rm conf}<0,
 \label{eq:accept}
\end{equation}
The standard error is $\widehat{\mathrm{SE}}_{\rm conf}=\mathrm{sd}(\{\bar\delta_s\})/\sqrt{S'}$. Let $n_{\rm src}$ be the number of relevant source tokens in the confirmation batch. For a suffix exchange, these are tokens whose source codeword is one of the two exchanged codewords. For a prefix exchange, these are tokens whose source codeword belongs to either exchanged group. Confirmation requires $n_{\rm src}\ge n_{\min}$ with $n_{\min}=8$. Candidates below this threshold are marked as undetermined and rejected.

The search uses a budget of $C_{\rm dec}=4.8\times10^5$ decoder calls over $R=4$ rounds. Each round allocates $C_{\rm dec}/8$ calls to prefix search and the same number to suffix search. After each round, we evaluate~\eqref{eq:risk} on $\Dset_{\rm dev}$. These evaluations are not included in $C_{\rm dec}$. We select the checkpoint with the lowest development objective and choose the earliest checkpoint if values are tied. The selected assignment is stored in two \texttt{uint16} lookup tables, each with $K$ entries. The search stops when the budget is exhausted or a complete round accepts no exchange. The procedure keeps $E$, $Q$, $D$, and $\C$ fixed, preserves the $LB$ payload bits, and leaves the reconstruction without channel errors unchanged.

With $K=MN$, hierarchical swapping has $\binom{M}{2}+M\binom{N}{2}=O(M^2+K^2/M)$ candidate exchanges, compared with $O(K^2)$ without grouping. For $K=16384$ and $M=32$, this gives about 32 times fewer candidates and reduces the decoder evaluations needed to test all candidates. CAC further considers only $O(q^2)$ suffix neighbors and tests at most $k=4$ candidates per sampled source. This smaller search space focuses the evaluation budget but limits the allowed assignments. It does not imply a speedup by a factor of 32 under a fixed budget of decoder calls.

\begin{algorithm}[t]
\caption{Budgeted Joint Index Reordering}
\label{alg:reorder}
\begin{algorithmic}[1]
\STATE \textbf{Input} $E,Q,D,\C$, $\Dset_{\rm tr},\Dset_{\rm dev}$, $\{\C_g\}$,
\STATE \hspace{1.1em} $\mathcal E,\{w_\varepsilon\}$, $C_{\rm dec},R$, $S,S',k,\kappa,n_{\min},\rho$
\STATE \textbf{Output} $\alpha,\{\beta_g\}$, lookup tables, diagnostics
\STATE build the geometric tree and set $\alpha_0,\beta_{0,g}$ to its paths
\STATE store $\mathbf z(x)$, estimate $\hat f$, and form $\mathcal A$
\FOR{$r=1,\dots,R$}
  \STATE spend $C_{\rm dec}/8$ calls on prefix search
  \STATE spend $C_{\rm dec}/8$ calls on suffix search
  \STATE evaluate \eqref{eq:risk} on $\Dset_{\rm dev}$ and save checkpoint
\ENDFOR
\STATE select the checkpoint with the lowest development objective
\STATE export the selected $\alpha,\{\beta_g\}$ and lookup tables
\STATE \textbf{procedure} Evaluate($\pi'$)
\STATE \quad screen on $S$ and reject if mean $\ge 0$
\STATE \quad confirm on $S'$ and count relevant source tokens $n_{\rm src}$
\STATE \quad accept if $n_{\rm src}\ge n_{\min}$ and \eqref{eq:accept} holds
\STATE \quad otherwise reject or record undetermined
\end{algorithmic}
\end{algorithm}

\begin{figure}[t]
\centering
\includegraphics[width=0.95\columnwidth]{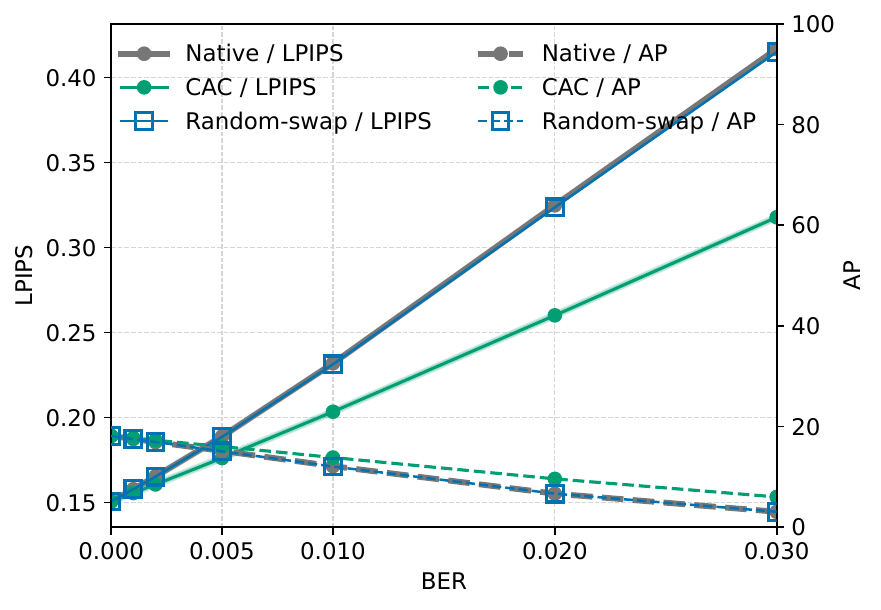}
\caption{LPIPS and COCO AP at different BERs for the native labels, the random swap baseline, and CAC.}
\label{fig:task}
\vspace{-1em}
\end{figure}

\begin{figure*}[t]
\centering
\includegraphics[width=\textwidth]{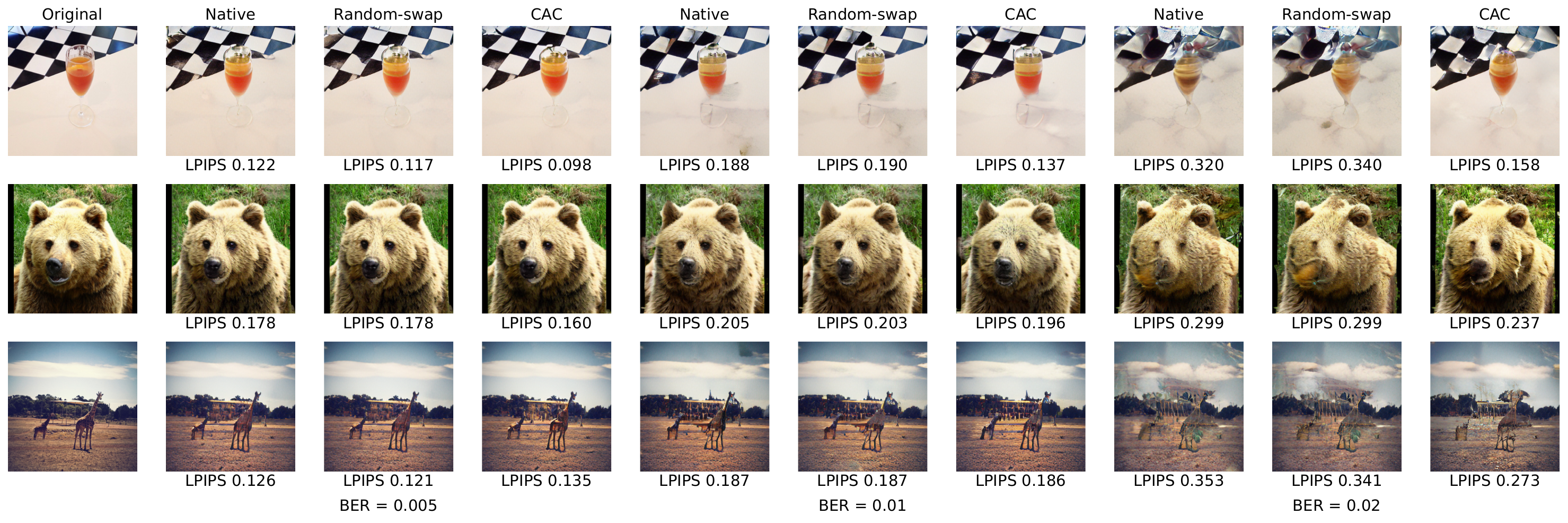}
\caption{Original images and reconstructions using the native labels, the random swap baseline, and CAC at different BERs.}
\label{fig:received-images}
\vspace{-1em}
\end{figure*}

% \begin{table}[!ht]
% \caption{Simulation Parameters}
% \label{tab:setup}
% \centering
% \small
% \setlength{\tabcolsep}{4pt}
% \begin{tabular}{@{}ll@{}}
% \hline
% Item & Setting\\
% \hline
% Visual tokenizer & VQGAN (ImageNet, f16), frozen\\
% Codebook & $K=16384$, $d=256$, $B=14$\\
% Hierarchy & $M=32$, $N=512$ ($p=5$, $q=9$)\\
% Canvas and payload & $256\times256$, $L=256$, $3584$ bits\\
% Training / development sets & $1024$ / $256$ ImageNet images\\
% Evaluation set & $2048$ COCO val2017 images\\
% BERs (train / test) & $0.001$, $0.005$, $0.01$ / $0$--$0.03$\\
% BER weights & equal over the design set\\
% Reference BER (paired tests) & $0.01$\\
% Search budget & $C_{\rm dec}=4.8\times10^{5}$, $R=4$\\
% Channel & BSC, ten realizations per image\\
% Reconstruction metric & AlexNet LPIPS v0.1\\
% Detection model & Faster R-CNN R50-FPN v2, frozen\\
% Detection metric & COCO AP, AP50, AP75\\
% \hline
% \end{tabular}
% \end{table}

\section{Simulation Results}
\label{sec:simulations}

\subsection{Simulation Setup}
All experiments use the fixed ImageNet f16 VQGAN described in Section~\ref{sec:model}. We set $M=32$, $N=512$, $p=5$, and $q=9$. This is an intermediate choice among the thirteen possible splits. Assignment training uses 1024 ImageNet images, and checkpoint selection uses a separate set of 256 ImageNet images. Final evaluation uses a fixed subset of 2048 COCO val2017 images~\cite{lin2014coco}, with its membership recorded by a hash. This subset does not overlap with the development set and is used for both reconstruction and detection evaluation. Images are resized with their aspect ratios preserved and padded to $256\times256$, giving $L=256$ tokens and a payload of 3584 bits. Reconstruction quality is measured using AlexNet LPIPS v0.1 relative to the preprocessed original image. Detection uses a fixed Faster R-CNN model with the ResNet-50-FPN v2 configuration pretrained on COCO. We report AP, AP50, and AP75. Training minimizes~\eqref{eq:risk} with equal weights at BERs of $0.001$, $0.005$, and $0.01$, using the budget and checkpoint selection rule in Section~\ref{sec:algorithm}. Evaluation applies independent BSC errors to all payload bits with ten noise realizations per image. All methods use the same noise masks.

\subsection{Reconstruction and Detection}

Fig.~\ref{fig:task} compares the test-set LPIPS and COCO AP of Native, Random-swap, and CAC with the visual backbone, detector, and payload unchanged. Random-swap starts from the native assignment and randomly exchanges codeword labels without using geometry or reconstruction loss, matching the number of codeword-pair exchanges induced by the accepted CAC updates. At $\varepsilon=0$, all three assignments achieve the same LPIPS of $0.1505$ and AP of $18.15$, consistent with bijective relabeling preserving error-free reconstruction.

Under channel errors, Random-swap remains close to Native, whereas CAC achieves lower average LPIPS and higher AP. At $\varepsilon=0.01$, Native, Random-swap, and CAC yield LPIPS values of $0.2318$, $0.2313$, and $0.2033$, and AP values of $12.14$, $12.07$, and $13.86$, respectively. CAC therefore reduces LPIPS by $12.3\%$ relative to Native and $12.1\%$ relative to Random-swap, while increasing AP by $1.72$ and $1.79$ points. At $\varepsilon=0.03$, CAC achieves an LPIPS of $0.3178$ and an AP of $6.02$, compared with $0.4168$ and $3.08$ for Native and $0.4150$ and $3.09$ for Random-swap. These results show that the tested random exchanges do not reproduce CAC's gains. Moreover, the AP improvement demonstrates a benefit to the evaluated downstream detection task, although detection accuracy is not included in the optimization objective.

Fig.~\ref{fig:received-images} presents three scenes reconstructed at BERs of $0.005$, $0.01$, and $0.02$ using matched channel realizations. At BER $0.02$, the LPIPS values for Native/Random-swap/CAC are $0.320/0.340/0.158$ for the glass, $0.299/0.299/0.237$ for the bear, and $0.353/0.341/0.273$ for the giraffes. CAC better preserves object contours and scene structure in these examples, whereas random exchanges provide no consistent improvement over Native. Individual reconstructions need not follow the average ordering: for the giraffe scene at BER $0.005$, CAC has a slightly higher LPIPS than both baselines. The visual results thus illustrate the average robustness gains without implying improvement for every image and noise realization.

\subsection{Comparison with Related Methods}
\begin{figure}[t]
\centering
\includegraphics[width=0.95\columnwidth]{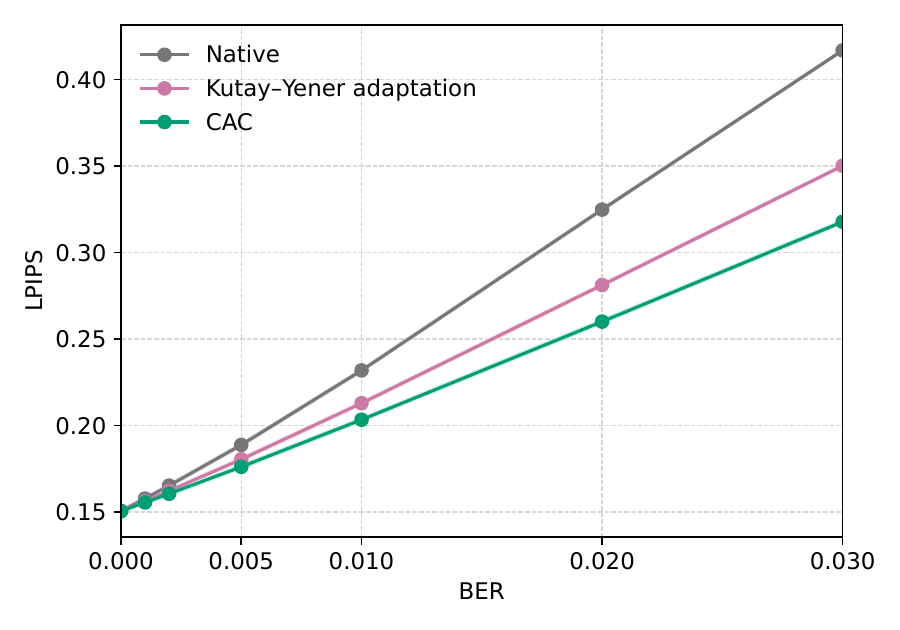}
\caption{LPIPS at different BERs for the native labels, the adapted method of Kutay and Yener, and CAC.}
\label{fig:literature}
\vspace{-1em}
\end{figure}

Fig.~\ref{fig:literature} compares Native, the assignment adaptation of Kutay and Yener~\cite{kutay2025}, and CAC at the same $3584$-bit payload. The adaptation applies geometry-supervised neural labeling followed by one-to-one assignment to the frozen VQGAN codewords. At $\varepsilon=0.01$, the three methods achieve LPIPS values of $0.2318$, $0.2129$, and $0.2033$, respectively, giving CAC a further $4.5\%$ reduction relative to the adaptation. CAC maintains a lower average LPIPS at every evaluated nonzero BER, including $0.3178$ versus $0.3501$ at $\varepsilon=0.03$.

Together with the Random-swap results in Fig.~\ref{fig:task}, this comparison shows that the complete CAC design outperforms both the tested random relabeling and the evaluated geometry-supervised assignment adaptation. Since Random-swap starts from Native rather than CAC's geometric initialization, it does not separately establish the contribution of the subsequent alternating optimization.

\section{Conclusion}

We proposed CAC to optimize frozen visual codebook labels for channel-induced perceptual distortion through hierarchical index reordering. Deployment uses static lookup tables, preserving the visual backbone, error-free reconstruction, and payload. At BER $0.01$, CAC reduces LPIPS by $12.3\%$ and improves AP by $1.72$ points over Native, while Random-swap remains close to Native. CAC also reduces LPIPS by $4.5\%$ relative to the evaluated geometry-supervised assignment adaptation. These results support codebook-label design for robust token communication without backbone retraining or additional payload bits. Future work will investigate grouping granularity and correlated channel errors.

\bibliographystyle{IEEEtran}
\bibliography{conference_101719}

@inproceedings{lin2014coco,
  author = {Tsung-Yi Lin and Michael Maire and Serge Belongie and James Hays and Pietro Perona and Deva Ramanan and Piotr Doll{\'a}r and C. Lawrence Zitnick},
  title = {Microsoft {COCO}: Common Objects in Context},
  booktitle = {Computer Vision---ECCV 2014},
  pages = {740--755},
  year = {2014},
  doi = {10.1007/978-3-319-10602-1_48}
}

@misc{liu2026tonic,
  author = {Sige Liu and Kezhi Wang},
  title = {{TONIC}: Token-Centric Semantic Communication for Task-Oriented Wireless Systems},
  year = {2026},
  howpublished = {arXiv:2605.21553},
}

@inproceedings{esser2021,
  author = {Patrick Esser and Robin Rombach and Bj{\"o}rn Ommer},
  title = {Taming Transformers for High-Resolution Image Synthesis},
  booktitle = {Proc. IEEE/CVF Conf. Computer Vision and Pattern Recognition (CVPR)},
  year = {2021},
  pages = {12873--12883},
}

@article{farvardin1990,
  author = {N. Farvardin},
  title = {A Study of Vector Quantization for Noisy Channels},
  journal = {IEEE Trans. Information Theory},
  volume = {36},
  number = {4},
  pages = {799--809},
  year = {1990},
  doi = {10.1109/18.53739}
}

@inproceedings{talbot1997,
  author = {N. L. C. Talbot and G. C. Cawley},
  title = {A Fast Index Assignment Method for Robust Vector Quantisation of Image Data},
  booktitle = {Proc. IEEE Int. Conf. Image Processing (ICIP)},
  volume = {3},
  pages = {674--677},
  year = {1997},
  doi = {10.1109/ICIP.1997.632211}
}

@article{knagenhjelm1996,
  author = {P. Knagenhjelm and E. Agrell},
  title = {The {Hadamard} Transform---A Tool for Index Assignment},
  journal = {IEEE Trans. Information Theory},
  volume = {42},
  number = {4},
  pages = {1139--1151},
  year = {1996},
  doi = {10.1109/18.508837}
}

@article{kutay2025,
  author = {Emrecan Kutay and Aylin Yener},
  title = {Harnessing the Power of Pre-Trained Models for Efficient Semantic Communication of Text and Images},
  journal = {Entropy},
  volume = {27},
  number = {8},
  pages = {813},
  year = {2025},
  doi = {10.3390/e27080813}
}

@inproceedings{zhang2018,
  author = {Richard Zhang and Phillip Isola and Alexei A. Efros and Eli Shechtman and Oliver Wang},
  title = {The Unreasonable Effectiveness of Deep Features as a Perceptual Metric},
  booktitle = {Proc. IEEE Conf. Computer Vision and Pattern Recognition (CVPR)},
  year = {2018},
  pages = {586--595},
}

@inproceedings{vandenoord2017,
  author = {Aaron van den Oord and Oriol Vinyals and Koray Kavukcuoglu},
  title = {Neural Discrete Representation Learning},
  booktitle = {Proc. Advances in Neural Information Processing Systems (NeurIPS)},
  year = {2017}
}

@book{gersho1992,
  author = {Allen Gersho and Robert M. Gray},
  title = {Vector Quantization and Signal Compression},
  publisher = {Kluwer Academic Publishers},
  address = {Boston, MA, USA},
  year = {1992},
  doi = {10.1007/978-1-4615-3626-0}
}

@article{kurka2020,
  author = {David Burth Kurka and Deniz G{\"u}nd{\"u}z},
  title = {{DeepJSCC-f}: Deep Joint Source-Channel Coding of Images With Feedback},
  journal = {IEEE J. Sel. Areas Inf. Theory},
  volume = {1},
  number = {1},
  pages = {178--193},
  year = {2020},
  doi = {10.1109/JSAIT.2020.2987203}
}

@article{gunduz2023,
  author = {Deniz G{\"u}nd{\"u}z and Zhijin Qin and Inaki Estella Aguerri and Harpreet S. Dhillon and Zhaohui Yang and Aylin Yener and Kai Kit Wong and Chan-Byoung Chae},
  title = {Beyond Transmitting Bits: Context, Semantics, and Task-Oriented Communications},
  journal = {IEEE J. Sel. Areas Commun.},
  volume = {41},
  number = {1},
  pages = {5--41},
  year = {2023},
  doi = {10.1109/JSAC.2022.3223408}
}

@book{cover2006,
  author = {Thomas M. Cover and Joy A. Thomas},
  title = {Elements of Information Theory},
  edition = {2nd},
  publisher = {Wiley},
  address = {Hoboken, NJ, USA},
  year = {2006}
}

@ARTICLE{11006980,
  author={Zhao, Zhouxiang and Yang, Zhaohui and Chen, Mingzhe and Zhu, Chen and Xu, Wei and Zhang, Zhaoyang and Huang, Kaibin},
  journal={IEEE Trans. Wireless Commun.}, 
  title={Energy-Efficient Probabilistic Semantic Communication Over Space-Air-Ground Integrated Networks}, 
  year={2025},
  volume={24},
  number={10},
  pages={8814-8829},
  doi={10.1109/TWC.2025.3569102}}

@misc{zhao2026fluidantennasystemsisac,
      title={Fluid Antenna Systems for ISAC Design: A Survey}, 
      author={Yueling Zhao and Ping Yang and Hong Niu and Shuaixin Yang and Yue Xiao and Gang Wu and Merouane Debbah and Ertugrul Basar},
      year={2026},
      eprint={2610.41382},
      archivePrefix={arXiv},
      primaryClass={eess.SP},
      url={https://arxiv.org/abs/2610.41382}, 
}

@article{xu2025energy,
  title={Energy efficient fluid antenna relay ({FAR})-assisted wireless communications},
  author={Xu, Ruopeng and Yang, Zhaohui and Zhang, Zhaoyang and Shikh-Bahaei, Mohammad and Huang, Kaibin and Niyato, Dusit},
  journal={IEEE Journal on Selected Areas in Communications},
  year={2025},
  publisher={IEEE}
}
\end{document}